# VZ Velorum: 116 years of a Mira star

**László L. Kiss**
*School of Physics, University of Sydney, NSW 2006, Australia*

**Abstract**    Using the Harvard College Observatory Photographic Plate Collection and recent CCD observations by the ASAS project we have reconstructed the light variations of the southern pulsating red giant star VZ Velorum between 1890 and early 2006. Contrary to an early report on its low-amplitude semiregular nature, we found a relatively stable Mira-like light curve with a mean period of 318 days and amplitude up to 7 magnitudes. The latest observations show evidence for a slightly shorter period (312 days). However, the difference does not exceed the intrinsic period jitter often seen in Mira type variables.

## 1. Introduction

Until recently, not much has been known about the southern variable star VZ Velorum. The data in the *General Catalogue of Variable Stars* (GCVS, Kholopov *et al.* 1985) originated from Payne (1928), who determined the period, epoch, maximum and minimum brightnesses from 76 Harvard plates taken in the last decade of the 19th and the first three decades of the 20th century. According to this report, VZ Vel changed semiregularly between 10.4 and 12.5 (photographic magnitudes) with a period of 317 days. Almost 50 years have passed until the next study, by Stephenson and Sanduleak (1976), who reported spectral types of southern long-period variables. They listed VZ Vel among stars which were not classified as Mira-type variables but nevertheless showing Mira-type spectra. The emission line spectrum and its IRAS Low Resolution Spectra (LRS) class also made it look more like a Mira than a semiregular variable (Kerschbaum and Hron 1996). And indeed, Lebzelter *et al.* (2005) obtained a radial velocity curve with shape and amplitude typical for Mira type stars. They also compared the data with the simultaneous ASAS *(All Sky Automated Survey)* V-band light curve (Pojmański 2002), showing that VZ Vel has to be classified as a Mira today. It was not clear, however, why the the star showed such a small amplitude in the observations made by Payne (1928).

This apparent contradiction raised our attention to the possible peculiar behavior of VZ Vel. Was it really a low-amplitude semiregular star in the early 20th century? Is it possible that there was a gradual or sudden increase in amplitude and/ or period? These are interesting questions because in the last decade we have learnt about several examples of the opposite behavior, namely changing from high-amplitude (Mira or Mira-like) to low-amplitude (semiregular) state, which is still a mystery. Examples include V Boo (Szatmáry *et al.* 1996, Pejcha and Greaves 2001), R Dor (Bedding *et al.* 1998), RU Cyg (Kiss *et al.* 2000), R Cen (Hawkins *et al.* 2001) and RU Vul (Templeton *et al.* 2005; Templeton 2006). Each of these stars underwent a strong amplitude decrease over decades, which was explained by different mechanisms, like beating between two close periods (Pejcha and Greaves 2001), possible mode switching with mildly chaotic behavior (Bedding *et al.* 1998), evolutionary change between Mira and semiregular states (Kiss *et al.* 2000) or effects of a He-shell flash (Hawkins *et al.* 2001). A few other Miras showed sudden amplitude declines, like Y Per (Kiss *et al.* 2000) and R V Cen (Mattei 1997), but those phases lasted only a few years, after which they rebounded (Hawkins *et al.* 2001).

Interestingly, no star has ever been reported with long-term amplitude increase resembling, for instance, the ~80 years of gradual decrease found in V Boo. This might be entirely due to observational selection effects, as low-amplitude variable stars have always been more difficult to discover and consequently, our observational records are heavily biased towards large-amplitude variables. The ultimate proof that it is indeed only a statistical bias would be the discovery of the first counter-example, which could be, for example, a strong argument for the beating model of Pejcha and Greaves (2001). That is why we decided to use the Harvard College Observatory Photographic Plate Collection to reconstruct light variations of VZ Vel during the 20th century.

## 2. The data

The brightness of VZ Vel was estimated on all available Harvard plates of the field with reference to a sequence of twelve nearby comparison stars covering uniformly the brightness range from B = 8.1 to B = 17.0. The magnitude values for the brighter reference stars were taken from *The Hipparcos and Tycho Catalogues* (Perryman *et al.* 1997), whereas for the fainter comparisons (B > 11.4) we used B magnitudes from *The Guide Star Catalogue* (GSC 2.2, STScI 2001). For a few stars present in both catalogues, the magnitude values differed only by 0.01–0.07 magnitudes, which is a reasonably good agreement considering that the primary purpose of the GSC 2.2 catalogue was to provide an accurate astrometric reference for the whole sky. In those cases we preferred using Tycho magnitudes.

The brightness estimates were made visually through a 9× viewing eyepiece with an estimated photometric accuracy of about ±0.1–0.2 mag. We found a noticeable non-linearity in the deepest plates, which made the faintest comparison stars barely distinguishable in brightness, so that our fainter magnitude estimates (B > 15) may have been affected by larger systematic error than the quoted uncertainties.

We inspected 831 blue plates in total, belonging to the RB/RH, MF, A, B, and Damon series. Of these, 528 yielded positive observations of VZ Vel, whereas all the other plates were not deep enough for a useful magnitude estimate. The earliest positive detection was on April 1, 1890, while the last one on June 30, 1989. The bulk of the observations were taken between the mid-1920s and early 1950s, when practically no light curve cycle was lost. We ignored the few yellow plates because they had very poor temporal coverage. Also, we did not take the upper-limit ("fainter than") observations into account, because the positive ones were enough to characterize the slow variations of the star. The full set of observations was deposited in the AAVSO International Database.

Finally, to supplement the historic light curve with the current variations, we also downloaded 295 CCD-*V* observations from the ASAS project. These data extend from late 2000 to early 2006 (http://archive.princeton.edu/~asas/).

## 3. Results

We show the full light curve in Figure 1. Apparently, VZ Vel has never been a low-amplitude semiregular star in the 20th century. All the available information show that it was and is a regular Mira star with cycle-to-cycle light curve changes up to 2 magnitudes in maximum brightness (between B = 10.0 and B = 12.0). The best Harvard plates showed that typical minima were fainter than B = 16.0, sometimes reaching close to B = 18.0. The late 19th century data did not have enough time coverage to allow measuring full light curve cycles, however, the overall magnitude distribution is consistent with the behavior in the 20th century.

We checked for the presence of multiple periodicity with Fourier analysis, using the PERIOD04 software of Lenz and Breger (2005). The power spectrum (Figure 2) revealed only one peak and its harmonics, with aliases, corresponding to a mean pulsation period of 317.97±0.03 days. This value is in perfect agreement with the 317-day period that was determined by Payne (1928), which suggests that her amplitude estimate was misled only by the bad sampling of the early data. Since the period is roughly halfway between 5/6 and 8/9 of a year, a poor temporal coverage, like the one in the upper panel in Figure 1, can disguise the full magnitude range even if the period can be determined accurately.

For the O–C diagram we combined the photographic and the CCD data. The epochs of maximum light were determined from polynomial fits in every case when both the ascending and descending branches were observed. Tentatively we also assigned epochs to the very early observations, when only single data points defined the times of maximum. We adopted ±15 days as probable error (i.e. 5% of the period). The resulting plot in Figure 3 suggests that recent ASAS maxima occur about 50 days (= 16% of the period) earlier than the predictions using the mean 20th century period and an epoch from 1988. This rather indicates a sudden change in period than the effects of the difference in photometric bands, because: (i) in Miras, like in other pulsating stars, the bluer the photometric band, the earlier the phase of the maximum light, so maxima in *V*-band occur later than in *B*; and (ii) other methods also suggest a slight change in the period of pulsations. Fourier analysis of the ASAS data alone yielded P = 311.8±0.3 days, while the phase diagrams confirm that recently, the period of pulsation is closer to 312 days than 318 days (Figure 4).

Similarly to the light curve, the given period change is also typical in a Mira star. The largest sample of Miras so far was studied by Templeton *et al.* (2005), who found period changes of a few percent over time-scales of 10–80 years in many Miras. The majority of stars showed random period "jitter," whereas in about 10% of the sample there was slow period "meandering." Templeton *et al.* (2005) speculated that the latter changes may represent thermal oscillations in the stellar envelope with characteristic Kelvin-Helmholtz timescales on the order of a few decades. What we find here for VZ Vel is consistent with the random period "jitter," although a recent onset of "meandering" cannot be excluded either.

In summary, we conclude that VZ Vel is a typical Mira star in all possible senses and the search for a pulsating red giant with gradual amplitude increase needs to be continued.

## 4. Conclusion

As Turner (2003) noted, the world's collection of archival photographic plate material comprises a wealth of useful information for the study of variable stars, yet it is often overlooked or underutilized. Large-scale CCD photometric survey projects like the *Northern Sky Variability Survey* (NSVS, Woźniak *et al.* 2004a) or the *All Sky Automated Survey* (ASAS, Pojmański 2002) yielded tens of thousands of new variable stars between V ~8–15 mag, with a high fraction of slowly varying red stars. For instance, Woźniak *et al.* (2004b) reported almost 8700 red giant variables from the 1-year observations of the NSVS project, of which approximately 1000 stars were newly identified Miras. On the other hand, the GCVS can be considered a complete sample of Mira stars only up to magnitude V = 9 (Kharchenko *et al.* 2002), so a huge number of today's Mira variables in the magnitude range that has already been covered by the photographic plate archives is yet to be discovered.

Obviously, to fully exploit the potentials of the photographic archives, the plates will have to be scanned and made publicly available. The digitization of the Harvard plate stacks is in a very early phase and a number of issues have yet to be solved, including the full astrometric and photometric calibrations of the plates (Mink 2006). Also, the combination of predominantly blue archive material and usually red/near-infrared CCD data may not be as straightforward as we hope. Nevertheless, the study of pulsating red giants will definitely be a major beneficiary of the digitization of photographic plate collections.

## 5. Acknowledgments

This work has been supported by a University of Sydney Postdoctoral Research Fellowship. The NASA ADS Abstract Service was used to access data and references. The author is grateful to Dr. Charles Alcock, the Director of the Harvard College Observatory, for access to the Photographic Plate Collection, and to Alison Doane, the Curator of the Collection, for helpful assistance during his visit to the archive.

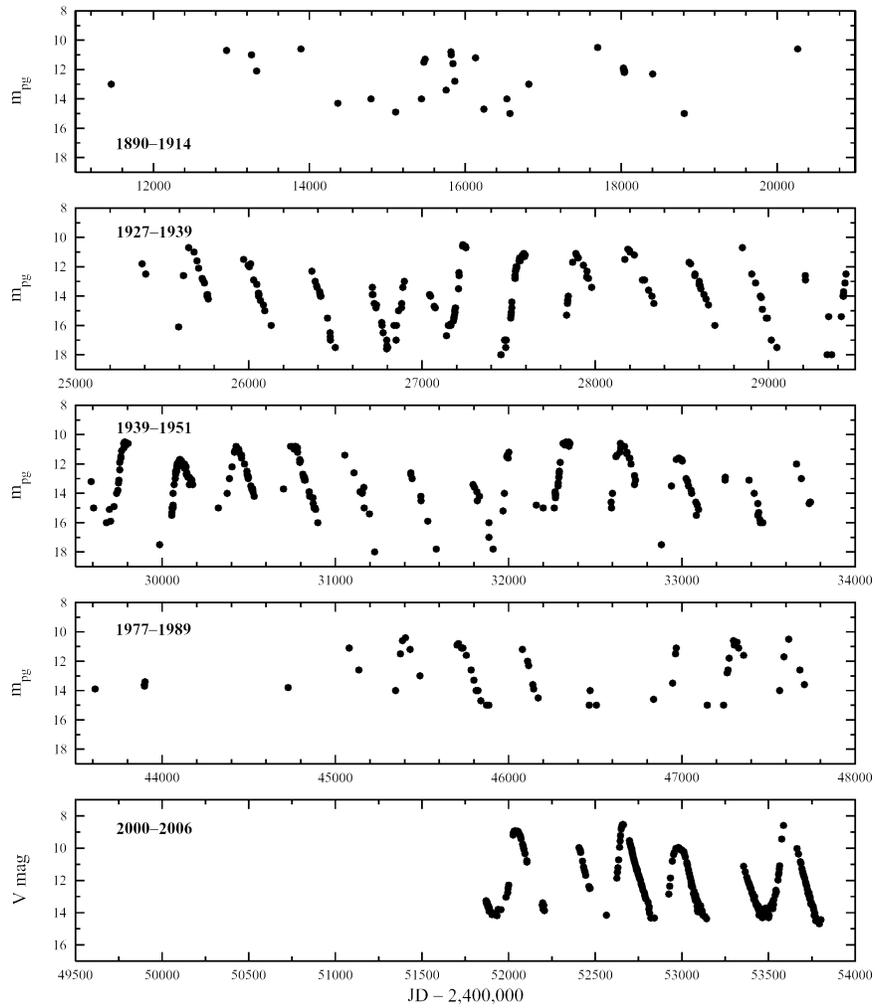

Figure 1. The light variations of VZ Vel between 1890 and early 2006. Note that the top panel shows twice as long a time interval as the other ones, while only the second and third panels from the top are continuous in time.

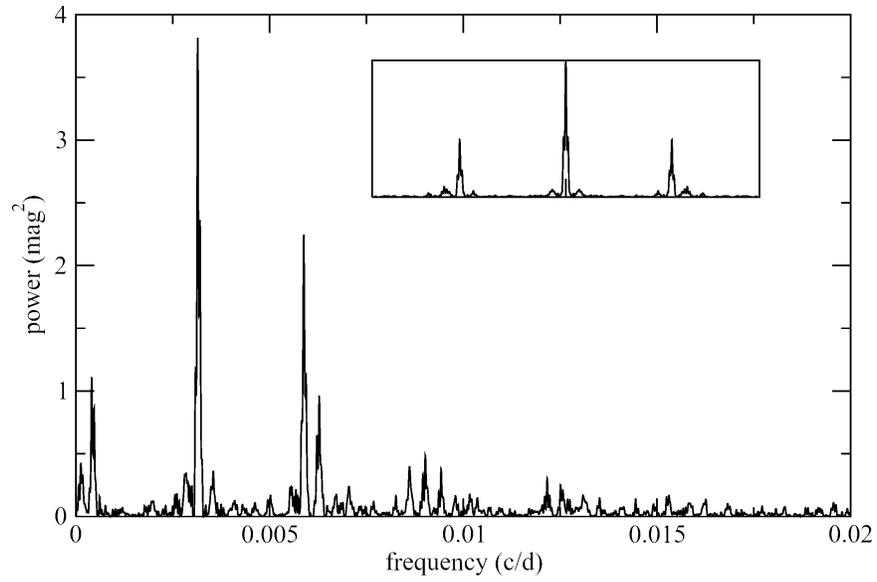

Figure 2. The power spectrum of the photographic light curve. The small insert shows the window function.

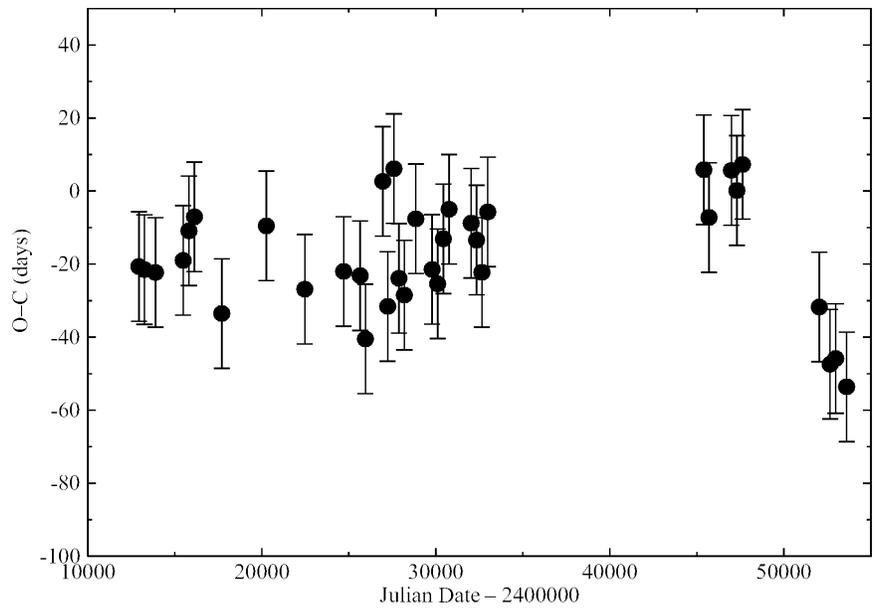

Figure 3. The O–C diagram of VZ Vel (P = 317.97d, E0 = 2447302).

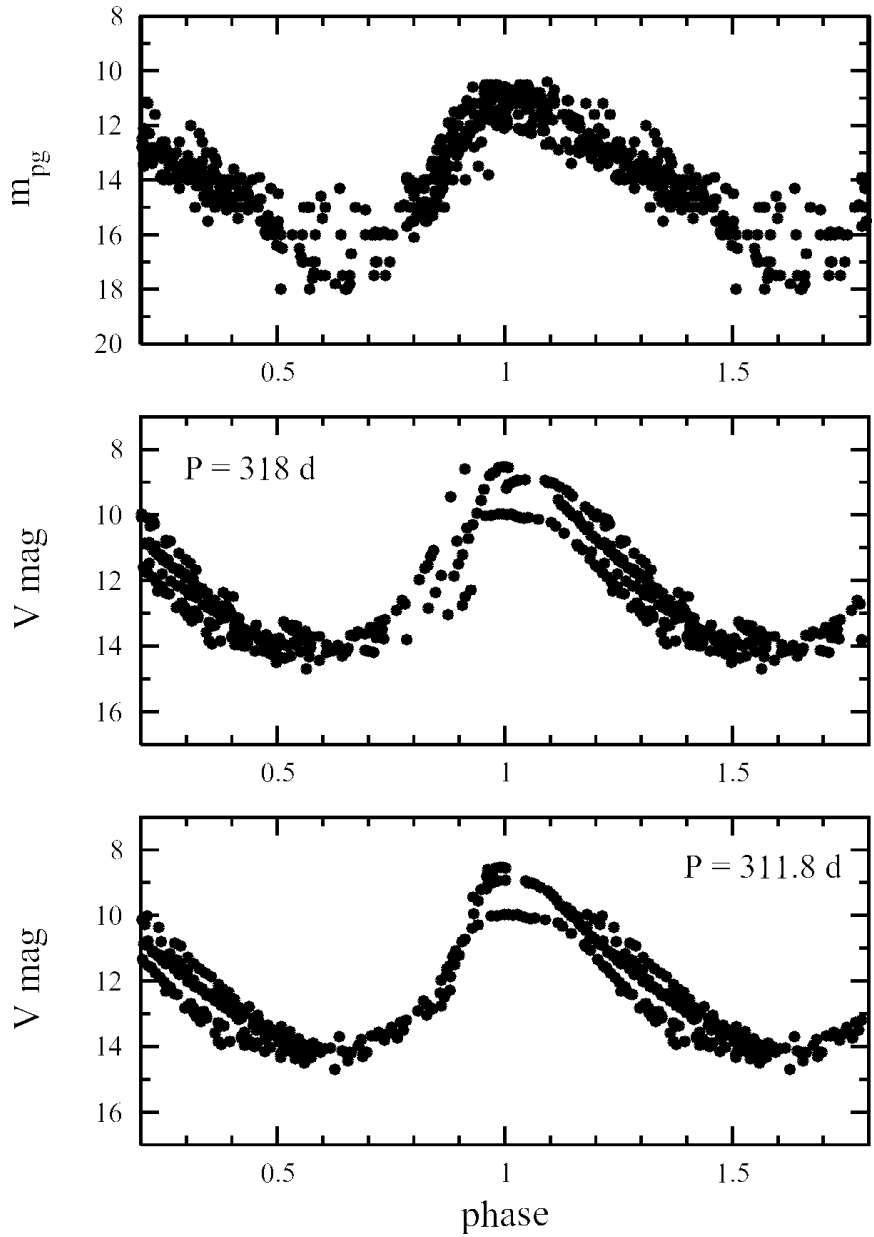

Figure 4. Phase diagrams of VZ Vel. Top panel: photographic data, P = 317.97d; middle panel: ASAS-3 V data, P = 318d; bottom panel: ASAS-3 V data, P = 311.8d. Note the decreased scatter of the ascending branch.